# A Model-based Approach for Effective Service Delivery

Feng-Lin Li, Chi-Hung Chi

Tsinghua University, Beijing, China

lifl08@mails.tsinghua.edu.cn

**Abstract.** With the prevalence of X-as-a-Service (e.g., software as a service, platform as a service, infrastructure as a service, etc.) and users' growing demand on good services, QoS (Quality of Service) assurance is becoming increasingly important to service delivery. Traditional service delivery mainly focuses on function or information provisioning, and does not give high priority to quality assurance. In this paper, we tackle the QoS assurance problem in a systematic way, from model to system. We first decompose traditional services into three components – namely software application, data and resource, then define models for these three kinds of basic services, and propose a set of operations for service publishing and composition. To illustrate our approach, we present a prototype system, the Platform as a Service (PaaS) system, which is developed in support of our framework and shows how QoS can be ensured through real-time monitoring and dynamic scaling up/down.

## 1. Introduction

In Service-Oriented Architecture (SOA), a service can be treated as a set of functionalities that can perform a specified task or provide certain information [1][2]. SOA thinks in terms of services and puts major emphasis on the standardization of service protocols and the interoperability among services. Through loosely coupled message exchanging, SOA allows to integrate services and applications conveniently at a minimum cost.

With the strong support of large companies such as Amazon, Microsoft and IBM, many kinds of services, e.g., Software as a Service (SaaS) [3], Platform as a Service (PaaS) [4] and Infrastructure as a Service (IaaS) [5], spring up quickly and become very popular in recent years. Also, the traditional triangular "Publish – Find – Bind" service delivery model [6] becomes service platform oriented. Under this new delivery model, service providers only need to upload their software and data on to a service platform, and can leave all the remaining affairs to the platform; users consume services through subscription in a pay-as-you-go manner.

In general, services are designed for offering certain functionality, information or capacity (e.g., computation resource). In addition to such functional properties, quality properties that indicate users' expectation for service behavior shall also be seriously considered. A case in point is the SLA compliance (conformity degree between the service runtime measurement and stated quality metrics in the Service Level Agreement) [7][8].

As a key aspect of services, the importance of QoS has been widely accepted. However, the related research mainly focuses on QoS estimation and aggregation in service selection and composition, but does not pay enough attention to quality assurance. For example, how can a user request be fulfilled in five seconds as stated in the SLA even at peak periods? How to satisfy an SLA (i.e, achieving good SLA compliance) with an economical cost?

In this paper, we focus on QoS assurance under the new service delivery paradigm. The remainder of the paper is structured as follows: Sec. 2 reviews the related work on service quality; Sec. 3 defines three kinds of basic services; Sec. 4 presents a service composition framework; Sec. 5 describes our prototype system – Platform as a Service (PaaS); Sec. 6 concludes the paper and sketch some interesting issues for future research.

## 2. Related Work

SOA is a paradigm for organizing and utilizing distributed capabilities to produce desired effects [10]. In this triangle delivery paradigm, a service is the representation of a repeatable business activity that has a specified outcome, and serves as a building block for composing larger services or applications. Meanwhile, a service shall be be self-contained, composable, autonomous, discoverable and location transparent [1]. Together with relevant implementation techniques such as Business Process Execution Language for Web Service (BPEL4WS) [11], Simple Object Access Protocol (SOAP) and Web Service Description Language (WSDL) [6], they have contributed strongly to the development of service industry.

However, under the new service delivery paradigm, some of these principles needed to be adapted. For example, if a service is self-contained, it will be harder to identify the problem or detect the bottleneck timely when a failure occurs. This is because a problem can be caused by different aspects, e.g., code defects, data inconsistency and network failure. In this case, if a service can be decomposed into proper components, things will become easier.

In general, services can be classified into two categories: *atomic* or *composite*. An atomic service is a service that is able to provide certain capability (information/functionality) independently. A *composite service r*elies on multiple services to finish a task, each of which can be either atomic or composite [12]. When constructing a composite service, one need to select proper component services and build a workflow out of them. The construction process of a composite service is called service composition [13]. Varieties of composition approaches, such as BPEL4WS [11], e-Flow [14], Star Web Services Composition Platform (StarWSCoP) [15], has been proposed. The majority of these techniques puts emphasis on service functionality rather than service quality.

Quality properties (or non-functional properties) specify how well a system shall behave. Typical quality properties include reliability, performance, availability and scalability [16]. As the importance of QoS is widely acknowledged, researchers have revised service models by adding QoS metrics [15][17]. These QoS metrics can be used for QoS calculation and estimation in service selection and composition [18][19][20]. On the other hand, there are also some useful tactics for improving service reliability and availability under the traditional delivery model: *active replication* (multiple instances of the same service will respond to a request and only a result will be selected), *passive replication* (another available instance of the same service would respond if the previous one failed to serve) [21][22], and *voting* (selecting the best response among the replicas) [23].

Although QoS calculation and aggregation are helpful for composing services with good theoretical quality metrics, they are ineffective and insufficient when it comes to QoS delivery at runtime. Also, the tactics aiming at improving reliability and availability also have limitations on performance, scalability and cost, since they do not support dynamic scaling up or down. For instance, even if a service has multiple instances on a platform, how about if all these instances are heavily overloaded? As another example, if a service has multiple instanc-es, can we remove some of them when the service is idle in order to reduce maintenance cost?

To sum up, a large body of work has been done on services, but it is still flawed on good QoS assurance. In this paper we propose models for three kinds of basic services, and a composition framework in support of QoS delivering. The key of our approach is dynamic scaling and load balance.

## 3. Basic Service Models

In SOA, a service is required to be self-contained and autonomous. Such self-contained properties become out of position in an open and collaborative service platform environment. For instance, an antivirus service may need to move to user-end for local data scanning; or an online picture editing service may need users to upload their pictures.

In this section, we define three types of basic services: *Data*, *Software* and *Resource service*. Each basic service model defines unified interfaces and proper properties in support of quality assurance based on real-time monitoring. Usually, interfaces specify how a service behaves in interactions and properties indicate what constraints it should follow.

### 3.1 Basic Data Service

In computer science, *data* is information represented in a form (usually binary) suitable for processing by computer [24][25]. It is often distinguished from program, which is a set of instructions that detail a task for the computer to perform. As a rule, it is replicable, movable and modifiable.

Nowadays, data indeed proves to be the 'Intel Inside' of the major applications and services. The great attractiveness of data is most likely due to the massive amount of potential information inside. Data itself carries no meaning. To generate value from data, we need to first collect data from different kinds of sources (e.g. flat file, XML, data table and stream), and then further interpret and process the collected data.

*Data service* are designed for the first purpose – data provisioning. A data service shall have an identifier and is able to provide required data effectively. As shown in Figure 1, there are three interfaces defined in our basic data service model.

- The *Definition Interface* declares service identifier, exchanging data format and data source. Usually, it also gives a clear description about data structure and its business value (usage or application).
- The *Application Interface* is designed for data provisioning and serves as a channel through which other services could obtain data. Note that it encapsulates the heterogeneity of different kinds of data sources.
- The *Management Interface* is reserved for platform operators to maintain. The manage operations include start, stop, resume and monitor that control service life cycles. Especially, *monitoring* plays a key role in QoS assurance: only through real-time monitoring, can we know service runtime status, detect bottlenecks and diagnose problems accurately. Furthermore, real-time monitoring also contributes to service intelligence like recommendation.

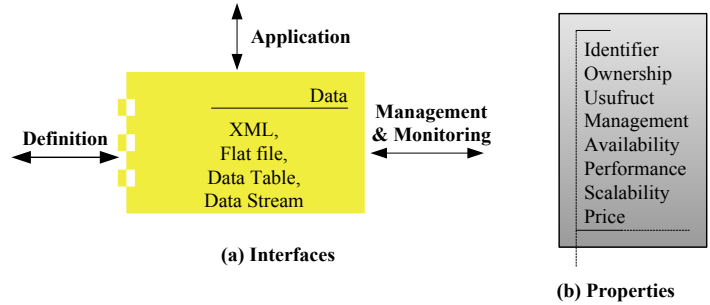

Figure 1 Basic Data Service Model

Having these interfaces, a data service can be discovered and invoked by other application software services. For service behavior, we need to specify constraints by introducing desired properties. Note that in a composition process, the properties of lower services would influence those of the resulting composite service. The properties of basic data service are shown in Table 1.

Table 1 Basic Data Service Properties

| Property | Description |
| --- | --- |
| *Identifier* | A unique symbol of service, usually specifies the access point |
| *Ownership* | Service builder (usually data owner or platform operator) |
| *Usufruct* | It is usually consumed by the upper software service |
| *Management* | Belongs to owner (usually delegated to platform operator) |
| *Availability* | In what period of time it can respond to user requests. |
| *Performance* | It usually implies how long it may take to respond |
| *Scalability* | How many parallel queries can it support at the same time? |
| *Price* | How much will an invocation or a subscription for a period costs |

## 3.2 Basic Application Software Service

*Application software*[1] is a computer program designed for people to perform a specific activity. It is contrasted with system software and middleware, which manage and integrate a computer's capabilities, but typically do not directly serve users [25]. Being similar to data, application software is replicable and movable. One point to note is that application code can be configurable through settings rather than modifying source code. For instance, it is possible to specify the access point of its data service in a software service's configuration file.

The basic software service model is defined as in Figure 2.

- The *Definition Interface* indicates service identifier, functionality, input and output. For example, the input of a java decompile service should be a standard *.class* file and the output will be a *.java* file which conforms to java language specification.
- The *Application* and *Invocation Interface*. When composing a composite service, an application software service has a dual identity. As a provider, it should exhibit its *Application Interface* to public for using; as a consumer, it needs to require other services via the *Invocation Interface* to perform an activity.
- The *Management* interface plays a similar role as that of data services. Also, m*onitoring* is the prerequisite of problem locating, SLA compliance measuring and QoS assurance. As for a composite service, monitoring can help us to find the most time-consuming sub-service or detect the problem accurately.

---

[1] We use the terms application, app, software, application software, and application code alternatively.

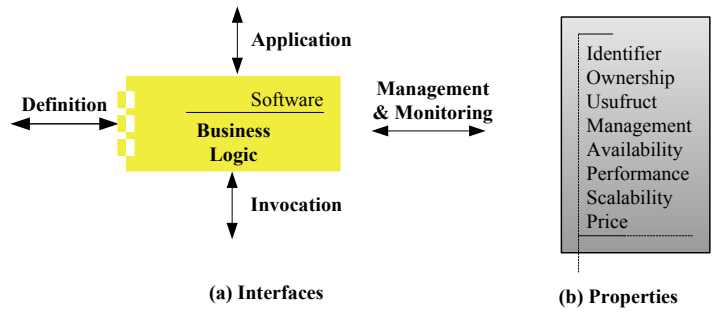

Figure 2 Basic Software Service Model

We also show the properties of application software services in Table 2.

Table 2 Basic Application Software Service Properties

| Property | Description |
| --- | --- |
| *Identifier* | A unique symbol and usually specifies the service access point |
| *Ownership* | It belongs to the person/organization that creates it via composition |
| *Usufruct* | Subscribers (end-users or upper software services) |
| *Management* | Owners usually delegate the right to platform operator |
| *Availability* | In what period of time the service keep work and respond |
| *Performance* | How long it may take to respond, usually an interval, e.g., 3~5s |
| *Scalability* | How many concurrent invocations it is able to support |
| *Price* | Users can pay per use or subscribe for a time period |

### 3.3 Resource as a service

Resources are physical or virtual entities that can be distributed in different place geographically. By resource, we refer to not only physical hardware such as CPU, memory and disk, but also software like operation system, service container and data base. For example, a virtual machine is often a combination of CPU, memory, disk, input/output, and operation system. Intuitively, resource differs from software or data mainly because it is un-replicable, unmovable and unmodifiable. Note that resource can be upgraded when needed, e.g., replacing the 512MB memory card with a 2GB one.

Resource service is able to deliver resource instance with specified capacity according to user requirements. As a new kind of service, resource service is quite different from the traditional software and data service. To provide resource service to consumers, the first step is to make physical resources available through Internet. In a subscription process, a consumer needs to negotiate with the resource provider at first, then subscribe resource with a specified configuration and finally obtain a resource instance with desired capacity. Based on the subscribed resource, consumer (vendors or platform operators) is able to compose services after certain environment configuration.

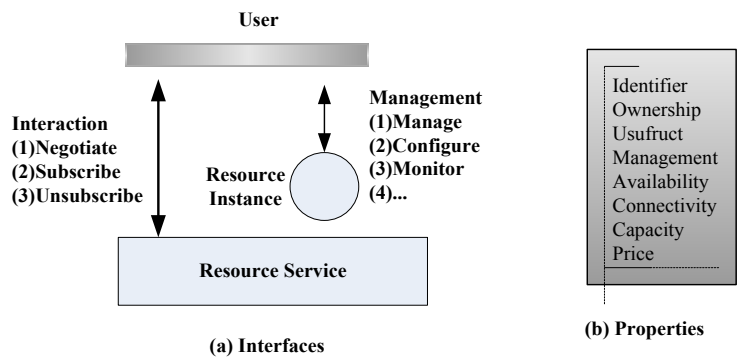

Figure 3 Resource as a Service Model

As shown in Figure 3, the resource service model contains two classes of operators. One class describes the interaction between users and providers while the other one tells how to use deliverable resource effectively.

- **Interactive Interface**
    (1) *Negotiate*. Agreement = Negotiate (User, Provider)

    When a user (often platform operator) is in need of resource, he/she at first need to achieve an agreement with resource providers about resource configuration, rental period, purpose/usage, authority, expense etc.

    (2) *Subscribe*. Resource Instance = Subscribe (User, Template, Agreement)

    Through subscription, a user is able to get the identify of a virtual machine that best matches his/her requirements stated in the submitted resource template [26]. With this identify, the user is able to use and manage the subscribed resource instance. This function can be overloaded to return a set of virtual resoruces.

    (3) *Unsubscribe*. Void = Unsubscribe (Resource Instance)

    A user can return back a rented resource and end his/her leasing agreement. In some cases, resource providers would also take back a resource if the leasing agreement expires or the user violates the achieved agreement (e.g., arrearage).

- **Management Interface**[2]
    (1) *Configure*. Void = Configure (Virtual Resource)

    A rented resource probably does not have desired environment configuration for software/data services when it is initially delivered. The configure operation allow users to install environmental software and configure system parameters.

    (2) *Monitor*. Stream Data = Monitor (Virtual Resource)

    This operation allows to monitor the runtime status and workload of resources. It collects data like usage of CPU, Memory, and Disk at certain frequency and output these data continuously. As such, monitoring provides real-time information for dynamic resource scaling.

Since consumers are more interested at how to use a resource instance effectively rather than how to get it, we put our emphasis on modeling delivered resource instances rather than static resource service. We show the properties of deliverable resource instances in Table 3 (as for those of resource service itself, readers can refer to those of software/data service).

---

[2] Unlike software/data service, users that subscribe resources only own the management right (start, stop and resume the resource instance), not the the ownership.

Table 3 Deliverable Resource Instance Properties

| Property | Description |
|---|---|
| *Identifier* | A unique symbol (e.g. an URL) that refers to a resource instance |
| *Ownership* | It still belongs to the resource provider |
| *Usufruct*[1] | Subscribers (usually platform operators) |
| *Management* | The lifecycle management is shifted to the subscribers |
| *Availability* | In what period of time the instance is active and can provide capacity |
| *Connectivity*[2] | It is able to connect with each other and exchange data |
| *Capacity*[3] | It is related with the resource fittings ( e.g. CPU, Memory, and Disk) |
| *Price* | It depends on the resource capacity, time period and operation system |

[1] It may have access control, such as process management, file operation and socket establishing
[2] E.g., Google App Engine has forbidden the socket and file operation, making it a solitary island
[3] It may include different classes, e.g. high throughput, large storage, computation centric instance

## 4. Service Publishing

In our framework, a service is a combination of software and/or data with resource. To publish services, we need to first turn software and data components into basic services, and then combine basic services according to certain rules to produce new services, which can be either atomic or composite. Note that basic services are different from atomic ones, which are self-contained and can finish tasks independently: an atomic service can be a basic service, or a combination of a basic software service and a basic data service.

### 4.1 Creating Basic Services

With our framework, building a basic service includes the following steps:

(1) Vendors *upload* software/data components to platform repository.
(2) Operators *negotiate* with resource providers to achieve an agreement.
(3) Obtaining a virtual instance via *subscription* with a specified configuration.
(4) Operators *start* the instance and *configure* system and software environment.
(5) Operators transmit the software with standard interfaces to the virtual instance by virtue of its *connectivity*, *deploy* and publish the component as a service.

Figure 4 Basic Service Building Steps

Through these five steps, we are able to get a basic service, of which the major properties depend on the software/data component and the underlying resource:
- The newly established basic service belongs to the component vendor/provider while its management right is delegated to platform operators.
- It can be used by end-users or invoked by other basic services.
- Its available time will not be longer than that of the resource instance.
- Its price may be affected by the resource pricing policy.
- Its performance is directly related with the resource capacity and visits.

### 4.2 Composing Atomic or Composite Services

In general, a composition of a basic software service and a basic data service mainly includes the four steps as in Figure 5. Through these steps, a software service is able to get data from a data service's application interface, and return specific result after processing.

> (1) Turn software and data component on the platform repository to basic services by following the instructions Figure 4.
> (2) (*Optional*) Set up a load balance and register the basic software service into it. Load balance will be in charge of distributing requests for load sharing.
> (3) (*Optional*) Set up a virtual data connector and register the basic data service to it. Virtual connector will work for data source selection and data provision.
> (4) Bind the software service and data service through configuring the data source in software service's invocation interfaces.

Figure 5 Service Composition between Basic Software and Data Service

The properties of a composed service depend on those of underlying services. For example, the composed service has a unique identifier, belongs to the person/organization (component provider or platform operator) who composed it, and has a pricing policy influenced by its underlying services. Taking timeliness as an example, its valid time span will not be longer than the intersection of the lower services' available time span.

When constructing composite services, the key is the information flow between component services: an upper service will perform extra operations on the information provided by a lower service (the former acts as a consumer, the latter acts as a provider). One point to be noted is that the upper input and the lower output shall conform to a common data structure.

### 4.3 Maintenance (Scaling)

For QoS assurance, real-time monitoring is of key importance. As workload increases, a single service instance may become insufficient to meet the promised QoS metrics, e.g., response time. In such a situation, dynamic scaling is a useful to relieve the pressure.

Thanks to the monitoring interface of each basic service, we are able to know the resource burden rate and detect the bottle neck of a composite service. This is helpful for decision making: e.g., whether it is necessary to subscribe another resource instance, which basic service in the composite service to replicate. If decided, platform operator can publish another service instance by following the instructions in Figure 4 and 5.

To support the dynamic scaling/replication, a load balance mechanism is needed for distributing user requests. On one hand, the load balance acts as a registry for software service instances on a platform; on the other hand, it serves as a router and redirects user requests to appropriate service instances. As for the data service layer, a similar virtual data connector is also desired. With load balance and virtual data connector mechanisms, we are able to perform dynamic service scaling automatically.

## 5. Prototype System – Platform as a Service (PaaS)

In this section, we introduce our PaaS system that is developed in support of the above models and composition procedures.

### 5.1 Architecture

As shown in Figure 6, our PaaS platform includes six building blocks: *center manager, load balance, data proxy, app module, monitor and resource publish*.

- **Center Manager** includes two projects: *WS_Platform Manager* and *Service Manager*. *WS_Platform Manager* is responsible for basic data, software service publishing and replication. It also maintains the software service mapping (a service identifier may be mapped to multiple instances), data service mapping as well as software-data service mapping information. The software service and data service mapping will be sent to the load balance and the virtual data connector respectively, such that they can distribute requests to share load.

*Service Manager* takes charge of component uploading, service publishing and subscription. Service publishing is in fact a composition, achieved by selecting appropriate data, application software and resource.

- **Load Balance** is the access point of services on the PaaS platform. Since the multiple instances of a service are often distributed to different resource instances, the *load balance* has to maintain a identifier-instances mapping table, which will be updated when a new software service instance is installed or an existing software service instance is replicated. When a request arrives, the *load balance* will pick up an appropriate URL from the mapping list according to system workload and perform redirection.

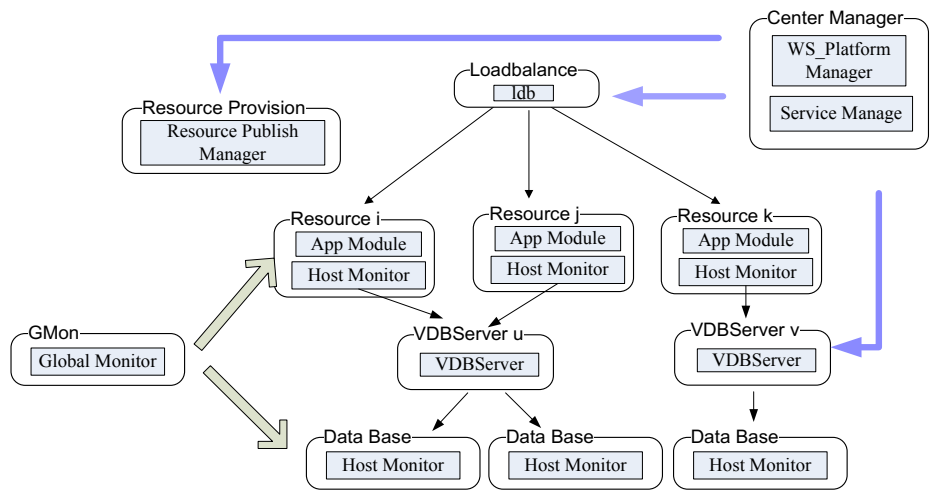

Figure 6 The PaaS Platform Structure

- **Data Proxy** (i.e. VDBServer) is responsible for data access. It contains the data mapping information which will be updated when a data service instance is published or replicated. In the mapping table, a list of data service instances is kept for one data service identifier. Data proxy also takes charge of data synchronization when a data service instance is replicated. This is achieved through coordination between data proxy and central manager. At last, it obtains data service runtime information through its monitoring interfaces and transmits to the global monitor timely.

- **App Module** takes charge of the deployment of application software. It needs to be installed on each resource instance that hosting software service. If a service is in need of replication, the app module is going to download the application code to the newly subscribed resource instance (with service runtime environment being configured) and publish it automatically. It will gather software service runtime information (e.g. response time) through their monitoring interfaces.

- **Monitor** contains two projects: *host monitor* and *global monitor*. *Host monitor* resides on each resource instance, memorizes the usage of CPU, Memory, Disk, Ethernet and Socket, put them into data streams and transfer to the global monitor for further analysis and decision making. It in fact acts like the monitoring interface of a resource in-stance. *Global Monitor* gathers information from app module, data proxy, and host monitor in the form of real-time data stream, and perform further processing. When workload is becoming heavy, the global monitor component is of key significance to detect the bottleneck (software or data part), decide whether to subscribe new resource instances or not. All of these will be done automatically.

- ***Resource Publish*** is a simple resource management system. In fact, a resource is often a physical or virtual entity, and only the meta information can be processed automatically. The component provides interfaces for obtaining idle and available resources according to user requirements.

**5.2 Illustration**

To better understand the service composition model proposed in section 4, here we are going to illustrate a typical service customization process, including creation, composition and scaling. In our system implementation, resource subscription, and service composition are of essential importance. In addition, the data consistency accompanying with dynamic data replication is also a key difficulty.

**5.2.1    Service Creation & Composition**

In our framework, software/data vendors only need to upload their software/data components that meet the format requirements of PaaS, and can leave the remaining affairs like service publish, maintenance (composition, scaling) to platform operators.

First, as shown in Eq.1, data/software providers package his/her component into a ZIP file by following the directory structure as in Figure 7 and upload it to the platform repository.

$$\text{Boolean} = \text{Upload (Provider, Software/Data Component)} \quad (1)$$

```
Software Component Package Compressed Format:
----Software_Example.zip //root
--------appcode
---------------META-INF
---------------WEB-INF
------------------------------lib
------------------------------web.xml
------------------------------classes
---------------------------------------*.classes
---------------index.jsp
Data Component Package Compressed Format:
----Data_Example.zip //root
--------data
---------------data.sql
The compress format should be ZIP and the archive should only contain monolayer root.
```

Figure 7 Platform software/data component archive format

Second, platform operators subscribe resource instances and publish basic services. As in Eq.2.1 ~ 2.2, when subscribing a resource instance, resource subscribers need to first achieve an Resource Service Level Agreement [27] with resource providers, which typically includes resource type, usage and pricing, and then request resources through resource template that states user requirements.

$$\text{Agreement} = \text{Negotiate (Platform Operator, Resource Providers)} \quad (2.1)$$
$$\text{Resource} = \text{Subscribe (Platform Operator , Resource Template, Agreement)} \quad (2.2)$$

In PaaS, we use the "*Resource* (*Value*)" pair [26][28][29] to indicate an entity and treat a resource as a combination of entities. For example, a resource template can be: "$R_v$ = {CPU (2GHZ, 2Core); Memory (1GB); Disk(40GB); OS(Ubuntu9.04); DB (MySQL5.01)}".

On receiving a subscription request, resource providers need to find a proper resource instance best matches user requirements through some matching algorithms. In general, a resource provider often offers several kinds of configurations, like small, large, and extra-large standard instances in the Amazon EC2 [30] platform. If the exiting resource provision does not match exactly to user requirements, the most similar resource instance will be returned.

For similarity measurement, a resource template is regarded as a vector and each attribute in the template is treated as a dimension. By doing so, we transform the matching problem to the calculation of distances between vectors, and use the weighted Euclidean Distance to perform the calculation. Note that this weighted algorithm allows assigning different weights to each attribute. For example, someone may require computation centric instance (assign larger weighting to CPU) while some others need high-memory instances.

Third, a delivered resource instance probably can not be directly put into use due to the lack of necessary environmental software. In this case, platform operators need to install necessary software and configure system parameters. If the resource instances are configured in advance by resource providers, then this step is optional. For example, in PaaS, a MySQL data base has been pre-installed for publishing basic data services, and a Tomcat/Jboss service container has been pre-installed for publishing basic application software services.

$$\text{Boolean} = \text{Configure (Resource Instance)} \quad (3)$$

The forth step is to publish basic services. At this step, platform operator needs to install the *App Module* and *Data Proxy* in the first place, and then deploy specified software service under the assistance of *App Module*. The *App Module* is responsible for downloading software package from platform repository and publishing it in the service container (Tomcat/Jboss) by invoking the hot deployment API of the container. Also, the data component will be installed into Mysql database and the data access interface is inserted into the data proxy mapping table. Moreover, if a data component is replicated, the consistency issue should be carefully considered. The important issue will be discussed in detail in dynamic scaling.

$$\text{Basic Service} = \text{Deploy (Platform Operator , Software/Data Component, Resource Instance)} \quad (4)$$

In PaaS, the publishing of services is supported by the *Service Manager* module. As shown in Figure 8, we are going to publish a data service on resource instance labeled with IP "192.168.10.99", then deploy a software service on on "192.168.10.100". When a basic service is published, it will inform the *WS_Platform Manager* module with a "(*Service: URL*)" pair indicates an available instance for a specified service. If it is a software service, *WS_Platform Manager* will inform *Load Balance* about the newly installed service instance. On receiving the new "(*Service: URL*)" pair, the load balance will perform duplicate checking: if the specified service already has instance(s) on the system, the load balance simply adds the new URL entry to the link list of that service; otherwise the balance will open a new link list and insert the "(*Service: URL*)" pair into it. As for basic data service, *WS_Platform Manager* will contact with the *Data Proxy* (*Virtual Data Connector*) to establish a data instance cluster. At last, *WS_Platform Manager* will write this informaiton into data base or file system.

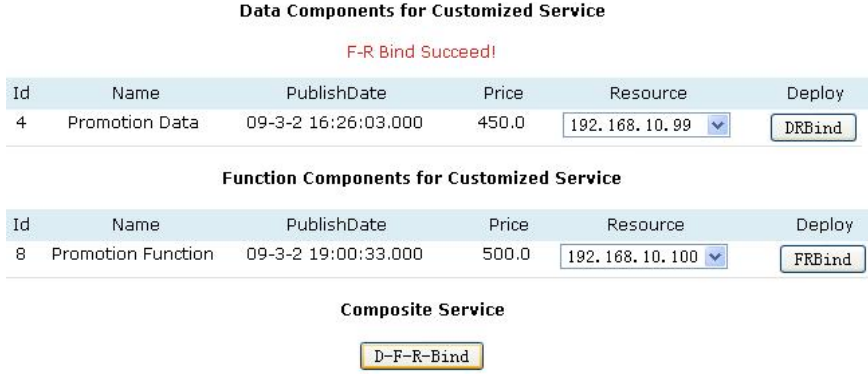

Figure 8 Service composition over basic services

Now we have published basic software/data services on PaaS. Since basic software/data services are often insufficient to fulfill a task or activity, we need to compose them in order to provide useful services. Eq.5 states the composition operator in our PaaS system, and Figure 8 shows how platform operators can perform the composition of a basic software service and a basic data service with just a click on the "D-F-R Bind" button. Once clicked, the mapping between software and data service will be sent to the *WS_Platform Manager* module and the two basic services are combined.

$$\text{Service} = \text{Compose (Service, Service)}^3 \tag{5}$$

On its first request, the software service will query *WS_Platform Manager* for its corresponding data service. With the obtained data service identifier and data proxy information, the software service will commit the data request " (*Data Proxy: Data Service Identifier: Data Request*)" to a specific data proxy. The invoked data proxy will look up its mapping list, select a proper data service instance and commit the data request. Finally, the result will go back the same way to the software service.

In order to balance the work load over multiple data service instances, we use a polling mechanism for data requests distribution. Considering that the data proxy itself can be a bottleneck, we use a data proxy layer (i.e., a set of virtual data connectors) instead of a singular virtual data connector. As such, a data proxy is responsible for only some of the software service and more than one proxy will be reserved for a software service.

### 5.2.2 Dynamic Scaling Up/Down

Dynamic scaling is an effective approach for QoS assurance. In the case of probable SLA violation (e.g., heavy workload), dynamic scaling can help to subscribe more resources and amortize workload through real-time monitoring and dynamic replication.

$$\text{Stream} = \text{Monitor (Software/Data/Resource Service Instance)} \tag{6}$$

In PaaS, *App Module* is responsible for monitoring the software service runtime behavior; *Data Proxy* records the data access time; *Host Monitor* gathers data about the resources usage info like CPU, Memory and Disk. Keep in mind that *Host Monitor* resides on each resource instance of the platform; *App Module* resides on software service resource instance; also there may be multiple Data Proxies. *App Module* and *Data Proxy* will generate a data tuple on each

---

  Note that PaaS also allows composing atomic/composite services for more complicated ones. This is why we use "Service" as the parameter type rather than "Basic Service".

user request while *Host Monitor* collect the resource instance usage information at a certain frequency. Each tuple confirms to a pre-defined schema and the monitoring result turns out to be a continuous data stream as time moves on. Take *Host Monitor* for example, suppose that the schema is "{Timestamp, CPU, Memory, Disk}", then a tuple in this data stream can be "{2011-3-16 15:21:23, 56%, 210MB, 25GB}".

We use BOREALIS[4] [31], an engine for processing distributed streams, and make necessary changes to integrate and analyze distributed data streams. The data streams generated by *App Module*, *Data Proxy* and *Host Monitor* are fed into the adapted BOREALIS system, and sent to the *Global Monitor* after data union, cleaning, associating and aggregation. On receiving the processed data stream, the *Global Monitor* module will perform further analysis like clustering and event detecting. For example, to detect if a resource instance is overloaded (e.g., 85% CPU usage lasting for 3 minutes). If yes, it will decide to subscribe a new resource instance with certain resource configuration.

As shown in Eq.7, dynamic scaling is in fact a composite process, it includes the following steps: *Subscribe*, *Configure* (optional if delivered resource is pre-configured), *Deploy*, *Compose*, and *Monitor*. When the *Global Monitor* decides to replicate a basic service, it will first subscribe a resource instance, and then configure the newly rented instance if needed, deploy another basic service using corresponding software/data component, and compose the newly deployed service instance with the original service instance.

$$Scaling = Subscribe; Configure; Deploy; Compose; Monitor \qquad (7)$$

One problem accompanying with replication is data inconsistency. When a data service needs to be replicated, if there is only one instance, reading requests can still be satisfied while writing requests will be blocked in a queue. At the end of replication, the cashed writing requests will be operated on all the data instances, including the newly replicated one. If there are more than one data instances on the platform, one of them will be selected as the copy source, the others will be responsible for data requests (note that writing requests will be performed on all other data instances). At the end, we will synchronize the source and destination data instances through extracting writing requests from log and execute them on the two.

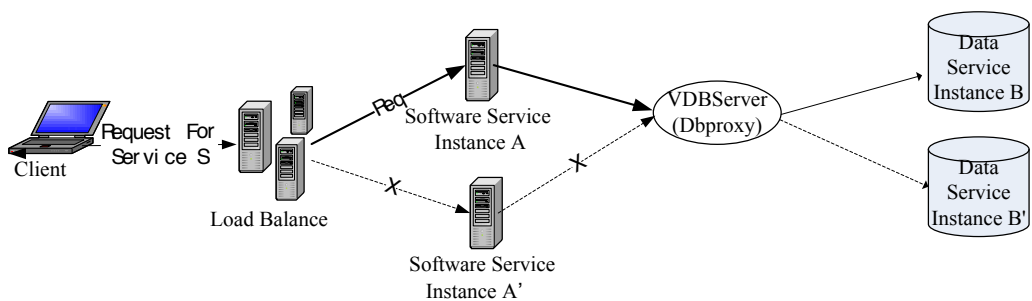

Figure 9 Request Process Flow

Suppose that both the basic software service A and the basic data service B are replicated as workload increases. As shown Figure 9, there are two software service instance A and A' while the data service instances are denoted as B and B'. A request arrives at the *Load Balance* first. Then the *Load Balance* will redirect it depends on the deployment structure and workload of the platform. Assuming the request is redirected to A in this case. And then A will send the request to the data proxy, who picks up a data source (B) and return back the result.

---

[4] There are several well-known streaming engines, such as STREAM [32], TelegraphCQ [33] and Aurora [34]. BOREALIS is designed for distributed application scenarios and fits well in our case while others are centralized.

## 6. Conclusion

In this paper, we tacked the QoS assurance problem in a systematic way: we first decompose traditional services into three components – namely software, data and resource, define models for these three kinds of basic services, and propose a set of operations for service publishing and composition. To illustrate our approach, we present a prototype system, the PaaS system, which shows how QoS can be ensured through dynamic scaling based on our service models and composition framework.

There are still many interesting issues remain open. For example, although we are able to scale up on-demand, we are not sure exactly how much extra capacity is in need to guarantee the desired QoS. If the issue can be worked out, it is very likely to attract service providers' interests, because of the potential cost savings.

Another interesting issue is how to detect the most resource-consuming software or data service instances with regarding to a resource instance. In our experiments, we have assumed that each resource instance holds only one basic service. Although we have the *App Module* (resp. *Data Proxy*) monitoring that gather information of each software (resp. data) instance, it is unable to get monitor how much CPU, Memory or Ethernet a software (resp. data) service instance has used. Our decomposition of services as software, data and resource allows to replicate components (software/data) of a service, it will be very interesting if we can identify which service instance on a resource node need to be replaced or removed.